\documentclass[twocolumn,aps,prl,preprintnumbers]{revtex4}
\usepackage{}
\usepackage{bbm}
\usepackage[latin9]{inputenc}
\usepackage{amsmath}
\usepackage{amssymb}
\usepackage{graphicx}
\usepackage{mathrsfs}
\usepackage{amsfonts}
\usepackage{amsthm}
\usepackage{color}
\usepackage{txfonts}
\usepackage{lipsum}
\usepackage[colorlinks=true,citecolor=blue,linkcolor=blue,urlcolor=blue,anchorcolor=blue]{hyperref}%
\hypersetup{colorlinks=true,citecolor=blue,linkcolor=blue,urlcolor=blue}
\providecommand{\U}[1]{\protect\rule{.1in}{.1in}}
\makeatletter
\@ifundefined{textcolor}{}
{
\definecolor{BLACK}{gray}{0}
\definecolor{WHITE}{gray}{1}
\definecolor{RED}{rgb}{1,0,0}
\definecolor{GREEN}{rgb}{0,1,0}
\definecolor{BLUE}{rgb}{0,0,1}
\definecolor{CYAN}{cmyk}{1,0,0,0}
\definecolor{MAGENTA}{cmyk}{0,1,0,0}
\definecolor{YELLOW}{cmyk}{0,0,1,0}
}
\makeatother
\begin{document}
\title{Realizing Corner States in Artificial Crystals Based on Topological Spin Textures}
\author{Z.-X. Li$^{1}$}
\author{Yunshan Cao$^{1}$}
\author{X. R. Wang$^{2,3}$}
\author{Peng Yan$^{1}$}
\email[Corresponding author: ]{yan@uestc.edu.cn}
\affiliation{$^{1}$School of Electronic Science and Engineering and State Key Laboratory of Electronic Thin Films and Integrated Devices, University of Electronic Science and Technology of China, Chengdu 610054, China}
\affiliation{$^{2}$Physics Department, The Hong Kong University of Science and Technology,
 Clear Water Bay, Kowloon, Hong Kong}
\affiliation{$^{3}$HKUST Shenzhen Research Institute, Shenzhen 518057, China}
\begin{abstract}
The recent discovery of higher-order topological insulators (HOTIs) has significantly extended our understanding of topological phases of matter. A conventional $n$-dimensional topological insulator (TI) has $(n-1)$-dimensional (first-order) topological edge states, whereas a higher-order, i.e., $k$th-order, where $2\leqslant k\leqslant n$, TI allows $(n-k)$-dimensional topological boundary modes, such as corner states and hinge states. These attractive features have promoted intensive research efforts in the community to observe novel states in a variety of nonmagnetic materials and structures. However, in the magnetism, even the second-order TIs are rarely reported. Here, we predict that second-order corner states can emerge in the dipolar-coupled dynamics of topological spin textures in two-dimensional artificial crystals. Taking a breathing honeycomb lattice of magnetic vortices as an example, we derive the full phase diagram of collective vortex gyrations and identify three types of corner states that have not been discovered before. We show that the topological ``zero-energy" corner modes are protected by a generalized chiral symmetry in the sexpartite lattice, leading to particular robustness against disorder and defects, although the conventional chiral symmetry of bipartite lattices is absent. We propose the use of the quantized $\mathbb{Z}_{6}$ Berry phase to characterize the nontrivial topology. Interestingly, we observe corner states at either obtuse-angled or acute-angled corners, depending on whether the lattice boundary has an armchair or zigzag shape. Full micromagnetic simulations confirm the theoretical predictions with good agreement. Experimentally, we suggest using the recently developed ultrafast Lorentz microscopy technique [M\"{o}ller \emph{et al}., \href{https://arxiv.org/abs/1907.04608}{arXiv:1907.04608}] to detect the topological corner states by tracking the nanometer-scale vortex orbits in a time-resolved manner. Our findings open up a promising route for realizing higher-order topologically protected corner states in magnetic systems and finally achieving topological spintronic memory and computing.
\end{abstract}

\maketitle
\section{INTRODUCTION}
Very recently, higher-order topological insulators (HOTIs) \cite{Benalcazar2017,Bernevig2017} have attracted extensive attention owing to their peculiar boundary modes at corners or hinges, which are absent from well-known first-order topological insulators (FOTIs) \cite{Hasan2010,Qi2011}. The topological description of HOTIs goes beyond the conventional bulk-boundary correspondence but is characterized by several new topological invariants, such as the nested Wilson loop \cite{Wan2017}, Green's function zeros \cite{Slager2015}, and the quantized bulk polarization (Wannier center) \cite{Ezawa2018,Song2017,Langbehn2017,Schindler2018}. Thus far, experimental evidence of HOTIs has been reported only in structured nonmagnetic materials \cite{Hassan2018,Mittal2018,Xue2019,Ni2019,Yang2019,Imhof2018,Serra2019,Serra2018,Kempkes2019}. To explore their applications in topological spintronics, it is intriguing to know whether magnetic HOTIs can exist from natural materials to metamaterials. An open question is what the proper topological invariant is for characterizing the higher-order topological phase in magnetic systems, which usually embody anisotropic long-range dipolar interactions.

The concept of topological insulator (TI) was originally conceived for electrons in solids. It is known that the perfect graphene lattice has a gapless band structure with Dirac cones in momentum space \cite{Castro2009}. When spatially periodic magnetic flux \cite{Haldane1988} or spin-orbit coupling \cite{Yao2007PRB} are introduced, a gap opens at the Dirac point, leading to a FOTI. Interestingly, the realization of gap opening and closing by tuning the intercellular and intracellular bond distances has been demonstrated in photonic \cite{Noh2018} and elastic \cite{Fan2019} honeycomb lattices. On the other hand, the topological phase of magnetic materials is of great current interest in magnetism because of its fundamental significance as well as in spintronics because of its practical utility for robust information processing \cite{Shindou2013,Mook2014,Chernyshev2016,Kim2016,Wang2017,Ruckriegel2018,Wang2018,Chen2018}. A particularly interesting system is magnetic texture-based metamaterials since they can offer flexible controllability that can benefit from modern spintronic techniques \cite{Vogel2010,Jung2011,Sugimoto2011,Han2013,Behncke2015,Adolff2016,Schulte2016,Velten2017,Im2017}. Indeed, topological chiral edge states have been predicted in a two-dimensional honeycomb lattice of magnetic textures (such as vortices and skyrmions) \cite{Kim2017,Li2018PRB}; however, these states are first-order by nature, according to the classification of topological insulators.

In this paper, we propose to achieve exotic HOTIs by designing artificial spin-texture crystals with alternating intercellular and intracellular bond lengths. We study dipolar-coupled magnetic vortices in a two-dimensional breathing honeycomb lattice without loss of generality. The collective motion of vortices is described by the generalized Thiele's equation, which manifests a wave-like equation in the artificial crystal and discriminates this equation from the wave equations of its electronic, photonic, and acoustic counterparts in the following aspects: (i) The nonvanishing topological charge induces a gyration term that is analogous to an effective magnetic field acting on a quasiparticle, thus breaking the time-reversal symmetry. (ii) The inertial effect is taken into account by a mass term. A third-order non-Newtonian gyration term is included to capture the high-frequency behavior of magnetic vortices and to determine the interaction parameters with high accuracy.\begin{figure}[ptbh]
\begin{centering}
\includegraphics[width=0.48\textwidth]{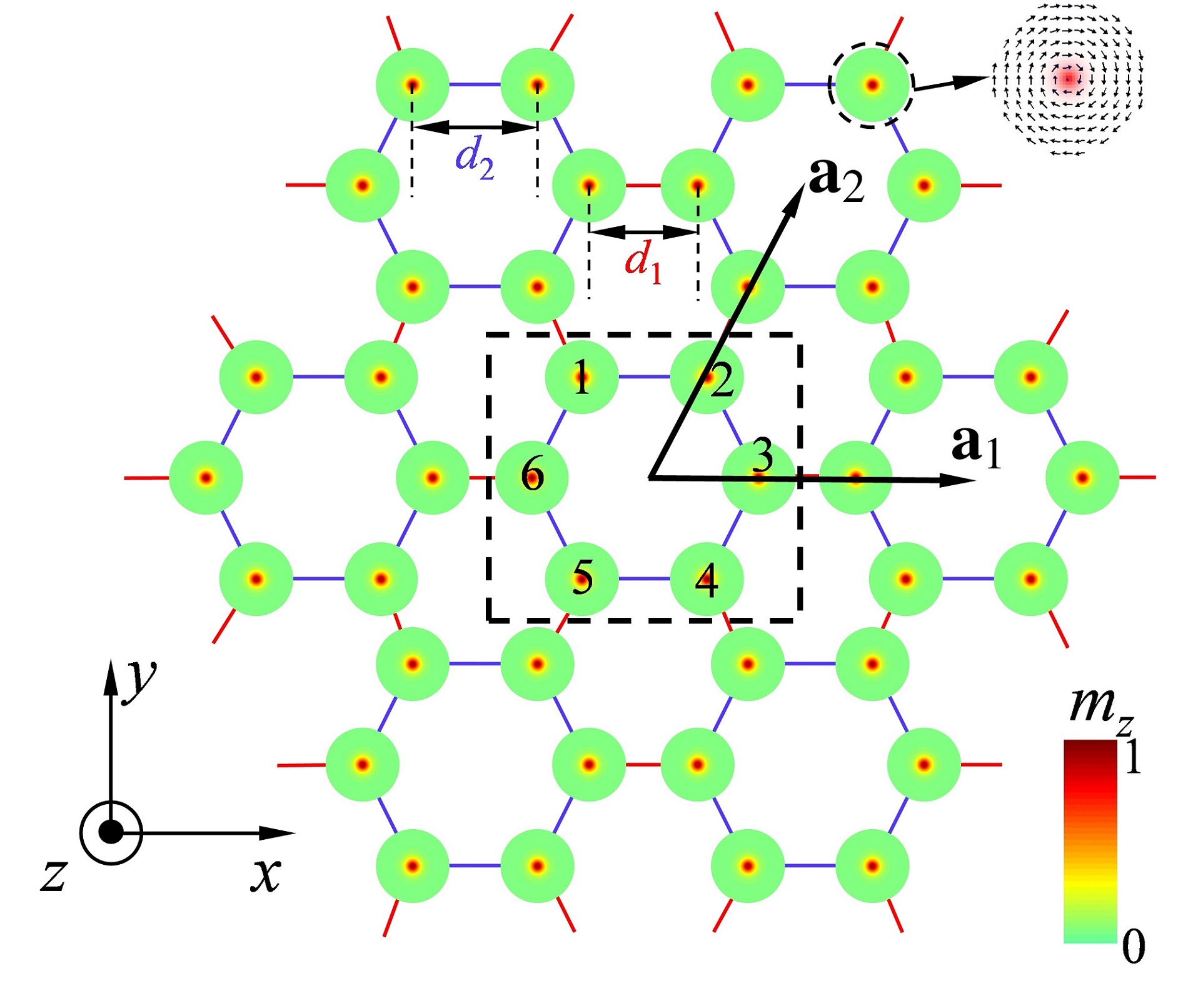}
\par\end{centering}
\caption{Illustration (top view) of the breathing honeycomb lattice of magnetic vortices, with $d_{1}$ and $d_{2}$ denoting the alternating lengths of intercellular and intracellular bonds, respectively. The radius of each nanodisk is $r=50$ nm, and the thickness is $w=10$ nm. The dashed black rectangle is the unit cell used for calculating the band structure, with $\mathbf{a}_1$ and $\mathbf{a}_2$ denoting the basis vectors.}
\label{Figure1}
\end{figure} (iii) The vortex-vortex coupling is strongly anisotropic. (iv) Conventional chiral symmetry in bipartite lattices is replaced by a more general chiral symmetry for the present sexpartite lattice. We prove that the generalized chiral symmetry protects a robust ``zero-energy" mode localized at lattice corners. By evaluating the Chern number and $\mathbb{Z}_{6}$ Berry phase of the bulk band structure, we obtain the full phase diagram and identify three topologically distinct phases that may exist in a finite lattice: (a) trivial phase without any topological edge state, (b) FOTI phase supporting the chiral edge mode localized at lattice boundaries, and (c) HOTI phase with topological corner states. These phases and their transition can be realized by tuning the ratio between intercellular and intracellular bond lengths ($d_{1}$ and $d_{2}$, respectively), as shown in Fig. \ref{Figure1}. The honeycomb lattice is called ``breathing" since $d_{1}\neq d_{2}$ in general. Furthermore, by modifying the overall shape of the lattice, we observe that topologically stable corner states emerge at obtuse-angled and acute-angled corners for armchair and zigzag edges, respectively, which is understood in terms of an intuitive physical picture in the zero-correlation limit. We also verify their topological stability against moderate disorder and defects. Finally, we implement full micromagnetic simulations for comparison with theoretical findings, reaching good agreement. Notably, the fabrication of artificial vortex or skyrmion lattices is readily within reach of current technology, e.g., electron-beam lithography \cite{Behncke2015,Han2013,Sun2013}. By tracking the nanometer-scale vortex orbits using the recently developed ultrafast Lorentz microscopy technique in a time-resolved manner \cite{Moller}, one can directly observe the topological corner states.
\quad\par
\section{MODEL, METHOD, AND BULK PROPERTIES}

We consider a breathing honeycomb lattice of magnetic nanodisks with vortex states, as shown in Fig. \ref{Figure1}. To describe the collective dynamics of the vortex lattice, we start with the generalized Thiele's equation \cite{Li2018PRB}
\begin{equation}\label{Eq1}
  \mathcal {G}_{3}\hat{z}\times\frac{d^{3}\textbf{U}_{j}}{dt^{3}}-\mathcal {M}\frac{d^{2}\textbf{U}_{j}}{dt^{2}}+\mathcal {G}\hat{z}\times \frac{d\textbf{U}_{j}}{dt}+\textbf{F}_{j}=0,
\end{equation}
where $\mathbf{U}_{j}= \mathbf R_{j} - \mathbf R_{j}^{0}$ is the displacement of the $j$-th vortex core (or guiding center) from the equilibrium position $\mathbf R_{j}^{0}$; $\mathcal {G}= -4\pi$$Qw M_{s}$/$\gamma$ is the gyroscopic constant, with $Q=\frac{1}{4\pi}\int \!\!\! \int{dxdy\mathbf{m}\cdot(\frac {\partial \mathbf{m}}{\partial {x} } \times \frac {\partial \mathbf{m}}{\partial y } )}$ being the topological charge [$Q=-1/2$ for the vortex configuration shown in Fig. \ref{Figure1}]; $\mathbf {m}$ is the unit vector along the local magnetization direction; $w$ is the thickness of the nanodisk; $M_{s}$ is the saturation magnetization; $\gamma$ is the gyromagnetic ratio; $\mathcal {M}$ is the effective mass of the magnetic vortex \cite{Makhfudz2012,Yang2018OE,Buttner2015}; and $\mathcal {G}_{3}$ is the third-order gyroscopic coefficient \cite{Mertens1997,Ivanov2010,Cherepov2012}. The conservative force can be expressed as $\textbf{F}_{j}=-\partial \mathcal {W} / \partial \mathbf U_{j}$, where $\mathcal{W}$ is the potential energy as a function of the vortex displacement: $\mathcal {W}=\sum_{j}\mathcal {K}\textbf{U}_{j}^{2}/2+\sum_{j\neq k}U_{jk}/2$, in which $U_{jk}=\mathcal {I}_{\parallel}U_{j}^{\parallel}U_{k}^{\parallel}-\mathcal {I}_{\perp}U_{j}^{\perp}U_{k}^{\perp}$ \cite{Shibata2003,Shibata2004}. Here, $\mathcal {K}$ is the spring constant, $U_{j}^{\parallel}$ ($U_{j}^{\perp}$) is the projection of the displacement $U_{j}$ onto (perpendicular to) the line connecting two vortices, and $\mathcal {I}_{\parallel}$ ($\mathcal {I}_{\perp}$) is the longitudinal (transverse) coupling constant.
By imposing $\mathbf{U}_{j}=(u_{j},v_{j})$ and setting $\psi_{j}=u_{j}+i v_{j}$, we have

\begin{equation}\label{Eq2}
   \hat{\mathcal {D}}\psi_{j}=\omega_{0}\psi_{j}+\sum_{k\in\langle j\rangle}(\zeta_{l}\psi_{k}+\xi_{l} e^{i2\theta_{jk}}\psi^{*}_{k}),
\end{equation}
where the differential operator $\hat{\mathcal {D}}=i\omega_{3}\frac{d^{3}}{dt^{3}}-\omega_{M}\frac{d^{2}}{dt^{2}}+i\frac{d}{dt}$, $\omega_{3}=\mathcal {G}_{3}/\mathcal {G}$, $\omega_{M}=\mathcal {M}/\mathcal {G} $, $\omega_{0}=\mathcal {K}/\mathcal {G} $, $\zeta_{l}=(\mathcal {I}_{\parallel, l}-\mathcal {I}_{\perp, l})/2\mathcal {G} $, and $\xi_{l}=(\mathcal {I}_{\parallel, l}+\mathcal {I}_{\perp, l})/2\mathcal {G}$, in which $l=1$ ($l=2$) represents the intercellular (intracellular) connection. $\theta_{jk}$ is the angle of the direction $\hat{e}_{jk}$ from the $x$-axis, where $\hat{e}_{jk}=(\mathbf{R}_{k}^{0}-\mathbf{R}_{j}^{0})/|\mathbf{R}_{k}^{0}-\mathbf{R}_{j}^{0}|$ and $\langle j\rangle$ is the set of nearest intercellular and intracellular neighbors of $j$. The key parameters $\mathcal {G}_{3}$, $\mathcal {M}$, $\mathcal {K}$, $\mathcal {I}_{\parallel}$, and $\mathcal {I}_{\perp}$ can be determined from micromagnetic simulations in a self-consistent manner \cite{Li2019}. For permalloy (Py) nanodisks, we have $\mathcal {G}=3.0725\times10^{-13}$ Js/m$^{2}$, $\mathcal{G}_{3}=4.5571\times10^{-35}$Js$^{3}$/m$^{2}$, $\mathcal {M}=9.1224\times10^{-25}$ kg, and $\mathcal {K}=1.8128\times10^{-3}$ J/m$^{2}$ \cite{Velten2017,Yoo2012}. The explicit expressions of $\mathcal{I}_{\parallel}(d)$ and $\mathcal{I}_{\perp}(d)$ as a function of $d$ are $\mathcal{I}_{\parallel}=M_{s}^{2}r\big(-1.72064\times10^{-4}+4.13166\times10^{-2}/d^{3}-0.24639/d^{5}+1.21066/d^{7}-1.81836/d^{9}\big)$ and $\mathcal{I}_{\perp}=M_{s}^{2}r\big(5.43158\times10^{-4}-4.34685\times10^{-2}/d^{3}+1.23778/d^{5}-6.48907/d^{7}+13.6422/d^{9}\big)$, which are obtained from the best fitting of numerical simulations of a two-body vortex-vortex interaction, where $d$ is the dimensionless distance between two vortices normalized by the nanodisk radius $r$.

We then rewrite the complex variable as
\begin{equation}\label{Eq3}
  \psi_{j}=\chi_{j}(t)\exp(-i\omega_{0}t)+\eta_{j}(t)\exp(i\omega_{0}t).
\end{equation}
For vortex gyrations, one can justify $|\chi_{j}|\gg|\eta_{j}|$. By substituting \eqref{Eq3} into \eqref{Eq2}, we obtain
\begin{widetext}
\begin{equation}\label{Eq4}
  \begin{aligned}
 \hat{\mathcal {D}}\psi_{j}=(\omega_{0}-\frac{\xi^{2}_{1}+2\xi^{2}_{2}}{2\omega_{K}})\psi_{j}+\zeta_{1}\sum_{k\in\langle j_{1}\rangle}\psi_{k}+\zeta_{2}\sum_{k\in\langle j_{2}\rangle}\psi_{k}
   -\frac{\xi_{1}\xi_{2}}{2\omega_{K}}\sum_{s\in\langle\langle j_{1}\rangle\rangle}e^{i2\bar{\theta}_{js}}\psi_{s}-\frac{\xi^{2}_{2}}{2\omega_{K}}\sum_{s\in\langle\langle j_{2}\rangle\rangle}e^{i2\bar{\theta}_{js}}\psi_{s},
  \end{aligned}
\end{equation}
\end{widetext}
where $\omega_{K}=\omega_{0}-\omega_{0}^{2}\omega_{M}$, $\bar{\theta}_{js}=\theta_{jk}-\theta_{ks}$ is the relative angle from the bond $k\rightarrow s$ to the bond $j\rightarrow k$ with $k$ between $j$ and $s$, and $\langle j_{1}\rangle$ and $\langle j_{2} \rangle$ ($\langle\langle j_{1}\rangle\rangle$ and $\langle\langle j_{2}\rangle\rangle$) are the set of nearest (next-nearest) intercellular and intracellular neighbors of $j$, respectively.

For an infinite lattice, the dashed black rectangle indicates the unit cell, as shown in Fig. \ref{Figure1}. $\textbf{a}_{1}=a\hat{x}$ and $\textbf{a}_{2}=\frac{1}{2}a\hat{x}+\frac{\sqrt{3}}{2}a\hat{y}$ are two basis vectors of the crystal, with $a=d_{1}+2d_{2}$. To calculate the band structure, we consider a plane wave expansion: $\psi_{j}=\phi_{j}\exp(-i\omega t)\exp\big[i(n\mathbf{k}\cdot\textbf{a}_{1}+m\textbf{k}\cdot\textbf{a}_{2})\big]$ where $j=1, 2, 3, 4, 5, 6$ denote different positions within the unit cell, as labeled in Fig. \ref{Figure1}; $n$ and $m$ are two integers; and $\textbf{k}$ is the wave vector. By substituting the above expression into Eq. \eqref{Eq4}, we obtain the matrix form of Hamiltonian in the momentum space

\begin{widetext}
\begin{equation}\label{Eq5}
 \mathcal {H}=\left(
 \begin{matrix}
   Q_{0} & \zeta_{2} & Q_{1}& \zeta_{1}\exp[i\textbf{k}\cdot(\textbf{a}_{2}-\textbf{a}_{1})]& Q_{2}& \zeta_{2} \\
   \zeta_{2} & Q_{0} & \zeta_{2}& Q_{3}& \zeta_{1}\exp(i\textbf{k}\cdot\textbf{a}_{2})& Q_{4} \\
   Q_{1}^{*} & \zeta_{2} & Q_{0}& \zeta_{2}& Q_{5}& \zeta_{1}\exp(i\textbf{k}\cdot\textbf{a}_{1})\\
   \zeta_{1}\exp[-i\textbf{k}\cdot(\textbf{a}_{2}-\textbf{a}_{1})] & Q_{3}^{*} & \zeta_{2}& Q_{0}& \zeta_{2}& Q_{6}\\
   Q_{2}^{*} & \zeta_{1}\exp(-i\textbf{k}\cdot\textbf{a}_{2}) & Q_{5}^{*} & \zeta_{2}& Q_{0}& \zeta_{2}\\
   \zeta_{2} & Q_{4}^{*} & \zeta_{1}\exp(-i\textbf{k}\cdot\textbf{a}_{1}) & Q_{6}^{*}& \zeta_{2}& Q_{0}
  \end{matrix}
  \right),
\end{equation}
with elements explicitly expressed as
\begin{equation}\label{Eq6}
\begin{aligned}
Q_{0}&=\omega_{0}-\frac{\xi_{1}^{2}+2\xi_{2}^{2}}{2\omega_{K}}, \\
Q_{1}&=-\frac{\xi_{1}\xi_{2}}{2\omega_{K}}\exp(i\frac{2\pi}{3})\Big\{\exp[i\textbf{k}\cdot(\textbf{a}_{2}-\textbf{a}_{1})]+\exp(-i\textbf{k}\cdot\textbf{a}_{1})\Big\}-\frac{\xi^{2}_{2}}{2\omega_{K}}\exp(i\frac{2\pi}{3}), \\
Q_{2}&=-\frac{\xi_{1}\xi_{2}}{2\omega_{K}}\exp(-i\frac{2\pi}{3})\Big\{\exp[i\textbf{k}\cdot(\textbf{a}_{2}-\textbf{a}_{1})]+\exp(i\textbf{k}\cdot\textbf{a}_{2})\Big\}-\frac{\xi^{2}_{2}}{2\omega_{K}}\exp(-i\frac{2\pi}{3}),\\
Q_{3}&=-\frac{\xi_{1}\xi_{2}}{2\omega_{K}}\exp(i\frac{2\pi}{3})\Big\{\exp[i\textbf{k}\cdot(\textbf{a}_{2}-\textbf{a}_{1})]+\exp(i\textbf{k}\cdot\textbf{a}_{2})\Big\}-\frac{\xi^{2}_{2}}{2\omega_{K}}\exp(i\frac{2\pi}{3}),\\
Q_{4}&=-\frac{\xi_{1}\xi_{2}}{2\omega_{K}}\exp(-i\frac{2\pi}{3})[\exp(i\textbf{k}\cdot\textbf{a}_{2})+\exp(i\textbf{k}\cdot\textbf{a}_{1})]-\frac{\xi^{2}_{2}}{2\omega_{K}}\exp(-i\frac{2\pi}{3}),\\
Q_{5}&=-\frac{\xi_{1}\xi_{2}}{2\omega_{K}}\exp(i\frac{2\pi}{3})[\exp(i\textbf{k}\cdot\textbf{a}_{2})+\exp(i\textbf{k}\cdot\textbf{a}_{1})]-\frac{\xi^{2}_{2}}{2\omega_{K}}\exp(i\frac{2\pi}{3}), \\
Q_{6}&=-\frac{\xi_{1}\xi_{2}}{2\omega_{K}}\exp(i\frac{2\pi}{3})\Big\{\exp[i\textbf{k}\cdot(\textbf{a}_{1}-\textbf{a}_{2})]+\exp(i\textbf{k}\cdot\textbf{a}_{1})\Big\}-\frac{\xi^{2}_{2}}{2\omega_{K}}\exp(i\frac{2\pi}{3}).
\end{aligned}
\end{equation}
\end{widetext}

\begin{figure}[ptbh]
\begin{centering}
\includegraphics[width=0.48\textwidth]{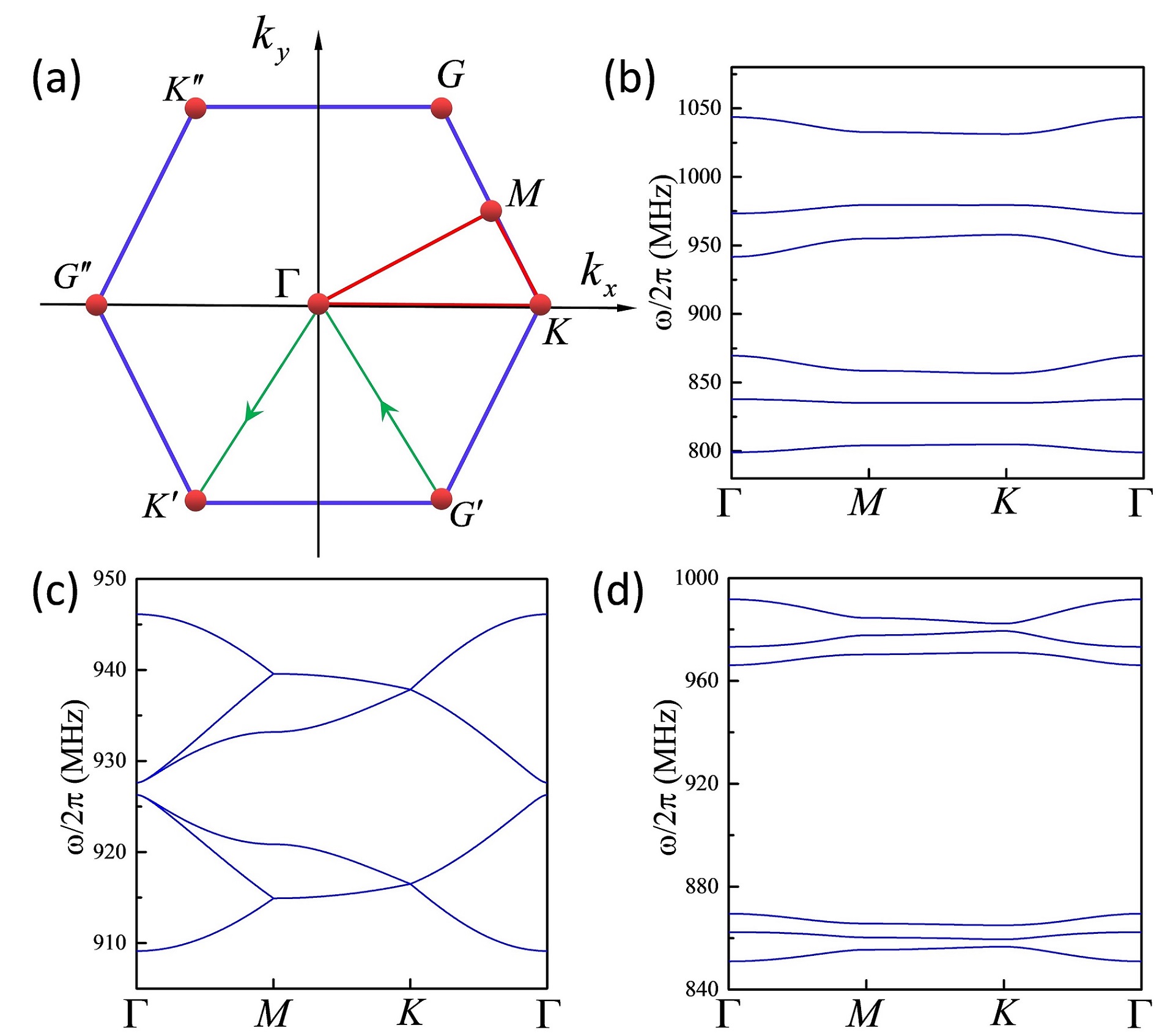}
\par\end{centering}
\caption{(a) The first Brillouin zone of the breathing honeycomb lattice, with high-symmetry points $\Gamma$, $G$, $M$, and $K$ located at $(k_{x},k_{y})=(0,0)$, $(\frac{2\pi}{3a},\frac{2\sqrt{3}\pi}{3a})$, $(\frac{\pi}{a},\frac{\sqrt{3}\pi}{3a})$, and $(\frac{4\pi}{3a},0)$, respectively. Band structures along the loop $\Gamma$-$M$-$K$-$\Gamma$ for different lattice parameters: $d_{1}=3.6r, d_{2}=2.08r$ (b), $d_{1}=d_{2}=3.6r$ (c), and $d_{1}=2.08r, d_{2}=3.6r$ (d).}
\label{Figure2}
\end{figure}
Topological invariants are useful to distinguish different phases, which are inherently determined by the system symmetry. For any insulator with translational symmetry, the gauge-invariant Chern number of bulk bands \cite{Wang2017,Avron1983}
\begin{equation}\label{Eq7}
   \mathcal{C}=\frac{i}{2\pi}\int\!\!\!\int_{\text{BZ}}dk_{x}dk_{y}\text{Tr}\big[P( \frac{\partial P}{\partial k_{x}}\frac{\partial P}{\partial k_{y}}- \frac{\partial P}{\partial k_{y}}\frac{\partial P}{\partial k_{x}})\big]
\end{equation}
is often adopted for determining the FOTI phase, where $P$ is the projection matrix $P(\textbf{k})=\phi (\textbf{k})\phi (\textbf{k})^{\dag}$, with $\phi (\textbf{k})$ being the normalized eigenstate (column vector) of \eqref{Eq5}, and the integral is over the first Brillouin zone. However, to determine whether the system allows the HOTI phase, we should consider a different topological invariant. To this end, we first analyze the symmetry of Hamiltonian \eqref{Eq5}. Because $(\xi_{1}^{2}+2\xi_{2}^{2})/2\omega_{K}\ll\omega_{0}$, the diagonal element of $\mathcal {H}$ can be regarded as a constant independent of $d$, i.e., $Q_{0}=\omega_{0}$, which is the ``zero-energy" of the original Hamiltonian. $Q_{1,2,3,4,5,6}$ are the next-nearest hopping terms. At first glance, the system does not possess any chiral symmetry to protect the ``zero-energy" modes because the breathing honeycomb lattice is not a bipartite lattice. Here, we generalize the chiral symmetry for a unit cell containing six sites by defining
\begin{equation}\label{Eq8}
\begin{aligned}
 \Gamma_{6}^{-1}\mathcal {H}_{1}\Gamma_{6}&=\mathcal {H}_{2},\\
 \Gamma_{6}^{-1}\mathcal {H}_{2}\Gamma_{6}&=\mathcal {H}_{3},\\
 \Gamma_{6}^{-1}\mathcal {H}_{3}\Gamma_{6}&=\mathcal {H}_{4},\\
 \Gamma_{6}^{-1}\mathcal {H}_{4}\Gamma_{6}&=\mathcal {H}_{5},\\
 \Gamma_{6}^{-1}\mathcal {H}_{5}\Gamma_{6}&=\mathcal {H}_{6},\\
 \mathcal {H}_{1}+\mathcal {H}_{2}+\mathcal {H}_{3}+\mathcal {H}_{4}&+\mathcal {H}_{5}+\mathcal {H}_{6}=0,
 \end{aligned}
\end{equation}
where the chiral operator $\Gamma_{6}$ is a diagonal matrix to be determined, and $\mathcal{H}_{1}=\mathcal{H}-Q_{0}\text{I}$. Upon combining the last equation with the previous five in Eqs. \eqref{Eq8}, we have $\Gamma_{6}^{-1}\mathcal {H}_{6}\Gamma_{6} =\mathcal {H}_{1}$, implying that $[\mathcal {H}_{1},\Gamma_{6}^{6}]=0$; thus, $\Gamma_{6}^{6}=\text{I}$, via the reasoning completely analogous to the Su-Schrieffer-Heeger model \cite{SSH1979}. Hamiltonians $\mathcal {H}_{1,2,3,4,5,6}$ each have the same eigenvalues $\lambda_{1,2,3,4,5,6}$. The eigenvalues of $\Gamma_{6}$ are given by $1,\ \exp(2\pi i/6),\ \exp(4\pi i/6),\ \exp(\pi i),\ \exp(8\pi i/6)$, and $\exp(10\pi i/6)$. Therefore, we can write
\begin{equation}\label{Eq9}
 \Gamma_{6}=\left(
 \begin{matrix}
   1 & 0 & 0& 0& 0& 0 \\
   0 & e^{\frac{2\pi i}{6}} & 0& 0& 0& 0 \\
   0 & 0 & e^{\frac{4\pi i}{6}}& 0& 0& 0\\
   0 & 0 & 0& e^{\pi i}& 0& 0\\
   0 & 0 & 0 & 0& e^{\frac{8\pi i}{6}}& 0\\
   0 & 0 & 0 & 0& 0& e^{\frac{10\pi i}{6}}
  \end{matrix}
  \right),
\end{equation}
in suitable bases. By taking the trace of the sixth line from Eqs. \eqref{Eq8}, we find $\sum_{i=1}^{6}\text{Tr}(\mathcal {H}_{i})=6\text{Tr}(\mathcal {H}_{1})=0$, which means that the sum of the six eigenvalues vanishes $\sum_{i=1}^{6}\lambda_{i}=0$. Given an eigenstate $\phi_{j}$ that has support in only sublattice $j$, it will satisfy $\mathcal {H}_{1}\phi_{j}=\lambda\phi_{j}$ and $\Gamma_{6}\phi_{j}=\exp[2\pi i(j-1)/6]\phi_{j}$ with $j=1,2,3,4,5,6$. From these formulas and Eqs. \eqref{Eq8}, we find that $\sum_{i=1}^{6}\mathcal {H}_{i}\phi_{j}=\sum_{i=1}^{6}\Gamma_{6}^{-(i-1)}\mathcal {H}_{i}\Gamma_{6}^{i-1}\phi_{j}=6\lambda\phi_{j}=0$, indicating $\lambda=0$ for any mode that has support in only one sublattice, i.e., zero-energy corner state.
In the presence of six-fold rotational ($C_{6}$) symmetry, a proper topological invariant is the $\mathbb{Z}_{6}$ Berry phase \cite{Zak1985,Kariyado2018,Hatsugai2011,Wakao2019,Mizoguchi2019,Araki2019}:
\begin{equation}\label{Eq10}
 \mathcal{\theta}=\int_{L_{1}}\text{Tr}[\textbf{A}(\textbf{k})]\cdot d\textbf{k}\ \  (\text{mod}\ 2\pi),
\end{equation}
where $\textbf{A}(\textbf{k})$ is the Berry connection:
\begin{equation}\label{Eq11}
 \textbf{A}(\textbf{k})=i\Psi^{\dag}(\textbf{k})\frac{\partial}{\partial\textbf{k}}\Psi(\textbf{k}).
\end{equation}
Here, $\Psi(\textbf{k})=[\phi_{1}(\textbf{k})$,$\phi_{2}(\textbf{k})$,$\phi_{3}(\textbf{k})]$ is the 6 $\times$ 3 matrix composed of the eigenvectors of Eq. \eqref{Eq5} for the lowest three bands. $L_{1}$ is an integral path in momentum space $G^{\prime}\rightarrow \Gamma\rightarrow K^{\prime}$; see the green line segment in Fig. \ref{Figure2}(a). The Wilson-loop approach is adopted for evaluating the Berry phase $\theta$ to avoid the difficulty of the gauge choice \cite{Benalcazar2017,Bernevig2017}. It is worth mentioning that the six high-symmetry points $G$, $K$, $G^{\prime}$, $K^{\prime}$, $G^{\prime\prime}$, and $K^{\prime\prime}$ in the first Brillouin zone are equivalent [see Fig. \ref{Figure2}(a)], because of the $C_{6}$ symmetry. Therefore, there are other five equivalent integral paths ($L_{2}: K^{\prime}\rightarrow \Gamma\rightarrow G^{\prime\prime}$, $L_{3}: G^{\prime\prime}\rightarrow \Gamma\rightarrow K^{\prime\prime}$, $L_{4}: K^{\prime\prime}\rightarrow \Gamma\rightarrow G$, $L_{5}: G\rightarrow \Gamma\rightarrow K$, and $L_{6}: K\rightarrow \Gamma\rightarrow G^{\prime}$) leading to the identical $\theta$. It is also straightforward to see that the integral along the path $L_{1}+L_{2}+L_{3}+L_{4}+L_{5}+L_{6}$ vanishes. Thus, the $\mathbb{Z}_{6}$ Berry phase must be quantized as $\theta=\frac{2n\pi}{6}\ $ $(n=0,1,2,3,4,5)$. By simultaneously quantifying the Chern number $\mathcal{C}$ and the $\mathbb{Z}_{6}$ Berry phase $\theta$, we can accurately determine the topological phases and their transition.

\section{CORNER STATES}
\subsection{Theoretical results}
To investigate the different phases of the system, we calculate the bulk band structures under a variety of lattice parameters, as shown in Figs. \ref{Figure2}(b)-\ref{Figure2}(d). For $d_{1}=d_{2}=3.6r$ [see Fig. \ref{Figure2}(c)], we find that the highest three bands and the lowest three bands merged separately, leaving a next-nearest hopping-induced gap centered at 927 MHz. In this case, the FOTI phase was anticipated \cite{Kim2017,Li2018PRB}. However, the six bands are separated from each other when we consider the other two kinds of parameters [see Figs. \ref{Figure2}(b) and \ref{Figure2}(d)], indicating that the system is in the insulating state. To further distinguish these insulating phases and the phase transition point, we examine the topological invariants Chern number and $\mathbb{Z}_{6}$ Berry phase of the lowest three bands below.

\begin{figure}[ptbh]
\begin{centering}
\includegraphics[width=0.48\textwidth]{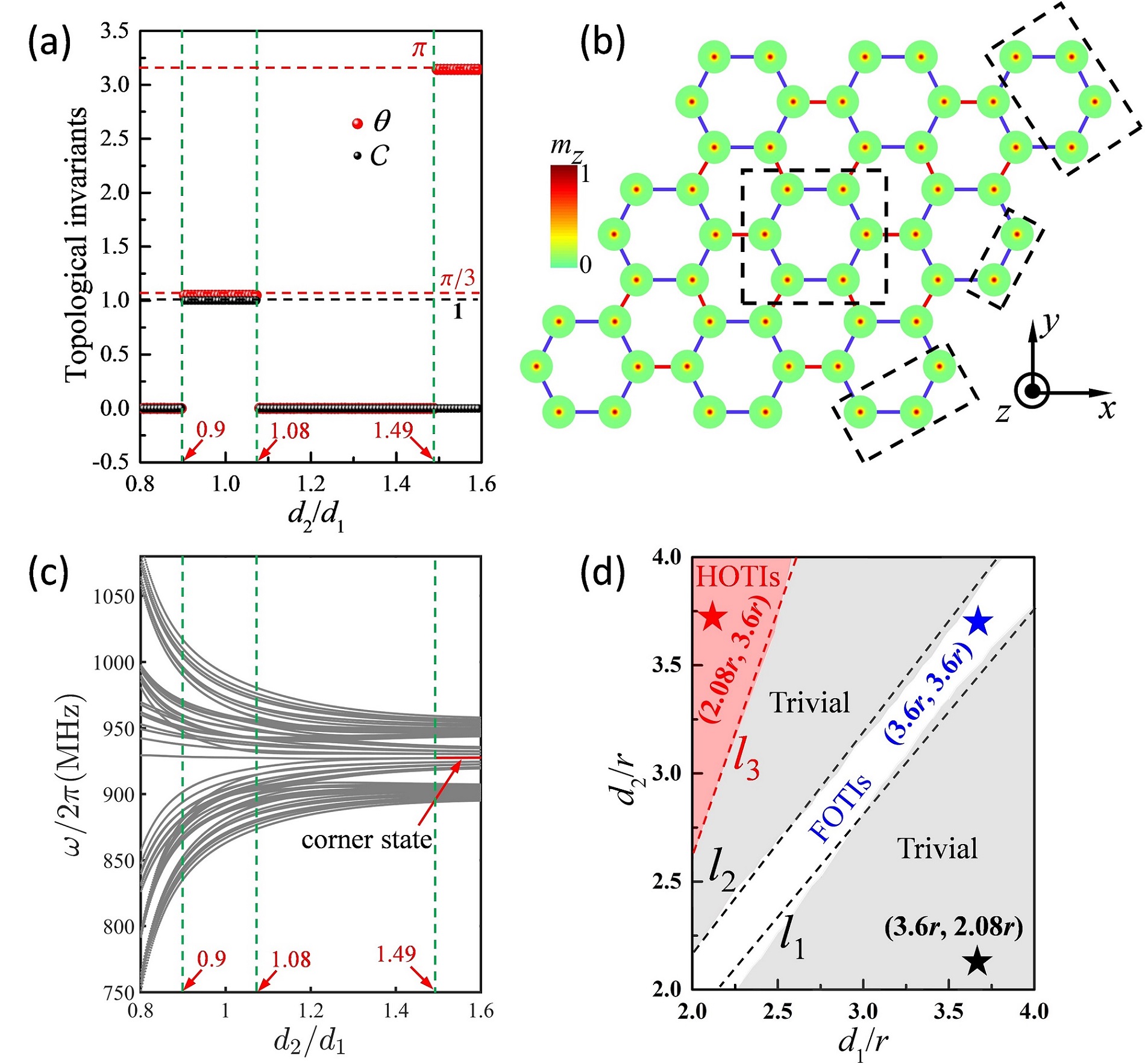}
\par\end{centering}
\caption{(a) Dependence of the topological invariants Chern number and $\mathbb{Z}_{6}$ Berry phase on the ratio $d_{2}/d_{1}$ when $d_{1}$ is fixed at $2.5r$. (b) Schematic plot of the parallelogram-shaped vortex lattice with armchair edges. (c) Eigenfrequencies of collective vortex gyration under different ratios $d_{2}/d_{1}$ with the red segment denoting the corner state phase. (d) Phase diagram of the system with pentagonal stars of different colors representing three typical parameters of $d_{1}$ and $d_{2}$ for different phases considered in the subsequent calculations and analyses.}
\label{Figure3}
\end{figure}

\begin{figure}[ptbh]
\begin{centering}
\includegraphics[width=0.48\textwidth]{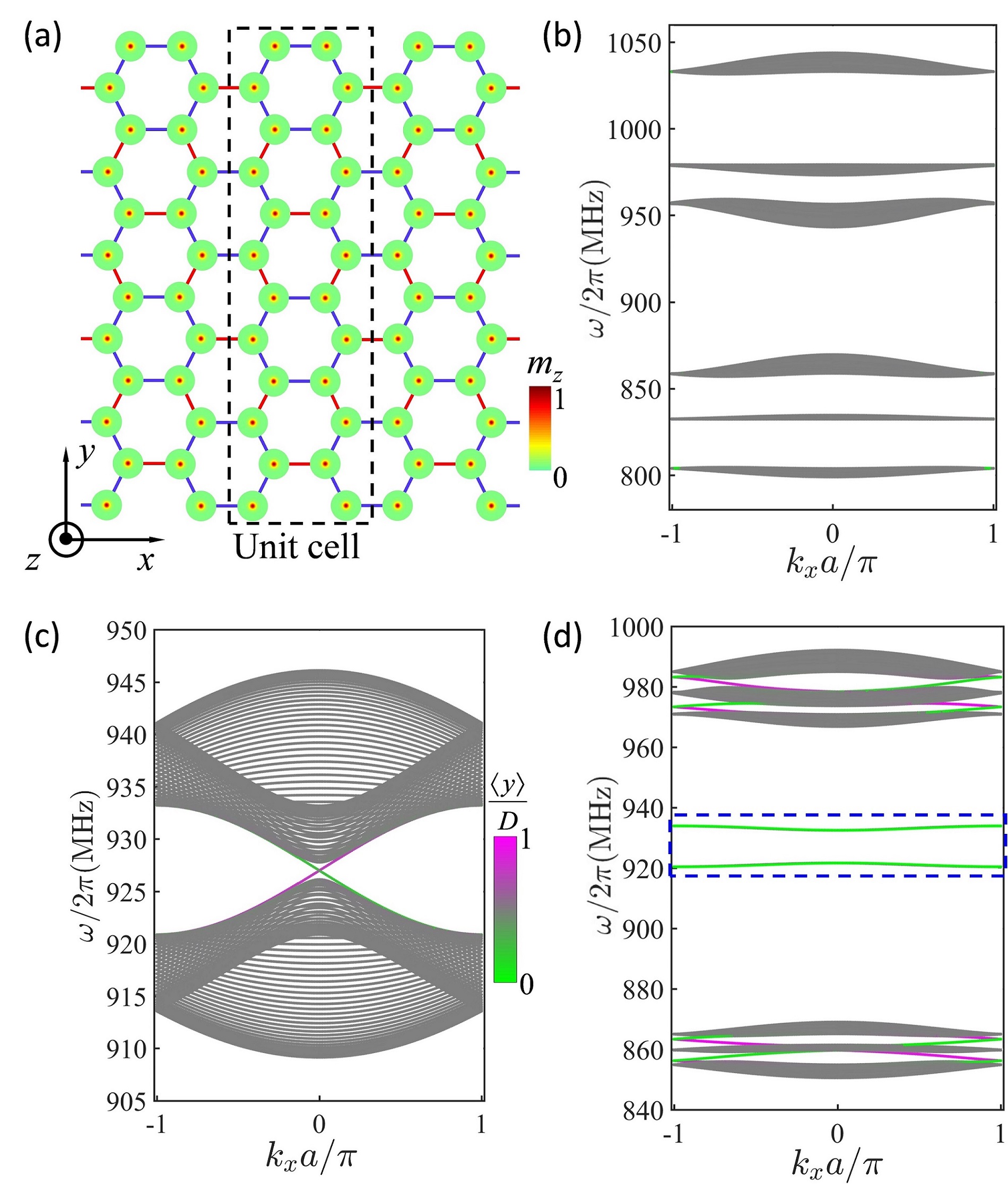}
\par\end{centering}
\caption{(a) Nanoribbon with armchair edges (closed boundaries in the $x$-direction and open boundaries in the $y$-direction). The dashed black rectangle is the unit cell. Band dispersions with different geometric parameters as denoted in Fig. \ref{Figure3}(d): $d_{1}=3.6r, d_{2}=2.08r$ (b), $d_{1}=d_{2}=3.6r$ (c), and $d_{1}=2.08r, d_{2}=3.6r$ (d). The dashed blue frame in (d) indicates the band of non-chiral edge states. $D$ is the width of the nanoribbon.}
\label{Figure4}
\end{figure}

Figure \ref{Figure3}(a) plots the dependence of the Chern number ($\mathcal{C}$) and the $\mathbb{Z}_{6}$ Berry phase ($\theta$) on the parameter $d_{2}/d_{1}$. In the calculations, we use the material parameters of Py (Ni$_{80}$Fe$_{20}$) \cite{Yoo2012,Velten2017} and fix $d_{1}=2.5r$. We can clearly see that $\theta$ is quantized to 0 when $d_{2}/d_{1}<0.9$ and $1.08<d_{2}/d_{1}<1.49$, to $\pi/3$ when $0.9<d_{2}/d_{1}<1.08$, and to $\pi$ when $d_{2}/d_{1}>1.49$, indicating that the system can support three different phases inside the gap between the 3rd and 4th bands. Furthermore, the Chern number is found to be +1 for $0.9<d_{2}/d_{1}<1.08$ and to vanish otherwise, showing that the system allows the FOTI phase. To clarify which phase the system is in when $d_{2}/d_{1}>1.49$, we calculate the eigenfrequencies of collective vortex gyration under different ratios $d_{2}/d_{1}$ for a parallelogram-shaped [see Fig. \ref{Figure3}(b)] structure, as shown in Fig. \ref{Figure3}(c). We observe that the system supports topologically stable corner states in the region $d_{2}/d_{1}>1.49$, which is the most obvious feature for the HOTI phase. We thus conclude that the system is in the trivial phase when $d_{2}/d_{1}<0.9$ and $1.08<d_{2}/d_{1}<1.49$, in the FOTI phase when $0.9<d_{2}/d_{1}<1.08$, and in the HOTI phase when $d_{2}/d_{1}>1.49$. To obtain the complete phase diagram, we systematically change $d_{1}$ and $d_{2}$ and evaluate topological invariants accordingly with the results plotted in Fig. \ref{Figure3}(d): The regions labeled gray, white, and red represent the trivial, FOTI, and HOTI phases, respectively. This diagram is the central result of the present work. Importantly, we find that the boundary for the phase transition between trivial and FOTI phases depends only weakly on the choice of the absolute values of $d_{1}$ and $d_{2}$ but is (almost) solely determined by their ratio, as indicated by dashed black lines ($l_{1}: d_{2}/d_{1}=0.94$ and $l_{2}: d_{2}/d_{1}=1.05$) in the figure. We further determine the boundary for the phase transition between trivial and HOTI phases by a linear function $l_{3}: d_{2}=2.24d_{1}-1.88$ [see the dashed red line in Fig. \ref{Figure3}(d)].

The existence of symmetry-protected states on boundaries is the hallmark of a topological insulating phase. To explicitly demonstrate the topological nature of the artificial crystal, we calculate the energy spectrum of the ribbon configuration with armchair edges [see Fig. \ref{Figure4}(a)]. By solving Eq. \eqref{Eq2} or Eq. \eqref{Eq4} numerically, we obtain the spectra of vortex gyrations with different choices of $d_{1}$ and $d_{2}$, as plotted in Figs. \ref{Figure4}(b)-\ref{Figure4}(d). For $d_{1}=3.6r$ and $d_{2}=2.08r$ [black star in Fig. \ref{Figure3}(d)], the system is in the trivial phase without any topological edge mode [see Fig. \ref{Figure4}(b)]. For $d_{1}=d_{2}=3.6r$ [blue star in Fig. \ref{Figure3}(d)], the lattice considered is identical to a magnetic texture version of graphene (the perfect honeycomb lattice). In contrast to the gapless band structure for perfect graphene nanoribbons, the imaginary second-nearest hopping term opens a gap at the Dirac point and supports a topologically protected first-order chiral edge state, consistent with previous findings \cite{Kim2017,Li2018PRB}. For $d_{1}=2.08r$ and $d_{2}=3.6r$ [red star in Fig. \ref{Figure3}(d)], one can clearly see two distinct edge bands, in addition to bulk ones, as shown in Fig. \ref{Figure4}(d). We confirm that these localized modes are actually not topological because they maintain the bidirectional propagation nature, which is justified by the fact that the wave group-velocity $d\omega/dk_{x}$ can be either positive or negative at different $k_{x}$ points. Below, we will show that higher-order topological corner states exactly emerge around these edge bands when the system is decreased to be finite in both dimensions. We also calculate the average vertical position of the modes $\langle{y}\rangle\equiv\sum_{j}R^{0}_{j,y}|\textbf{U}_{j}|^2/\sum_{j}|\textbf{U}_{j}|^2$ highlighted in the band structure, where $R^{0}_{j,y}$ is the equilibrium position of vortices projected onto the $y$ axis (a definition is given in Sec. IIIB), represented by different colors: Colors closer to magenta indicate modes more localized at the upper edge.

\begin{figure}[ptbh]
\begin{centering}
\includegraphics[width=0.48\textwidth]{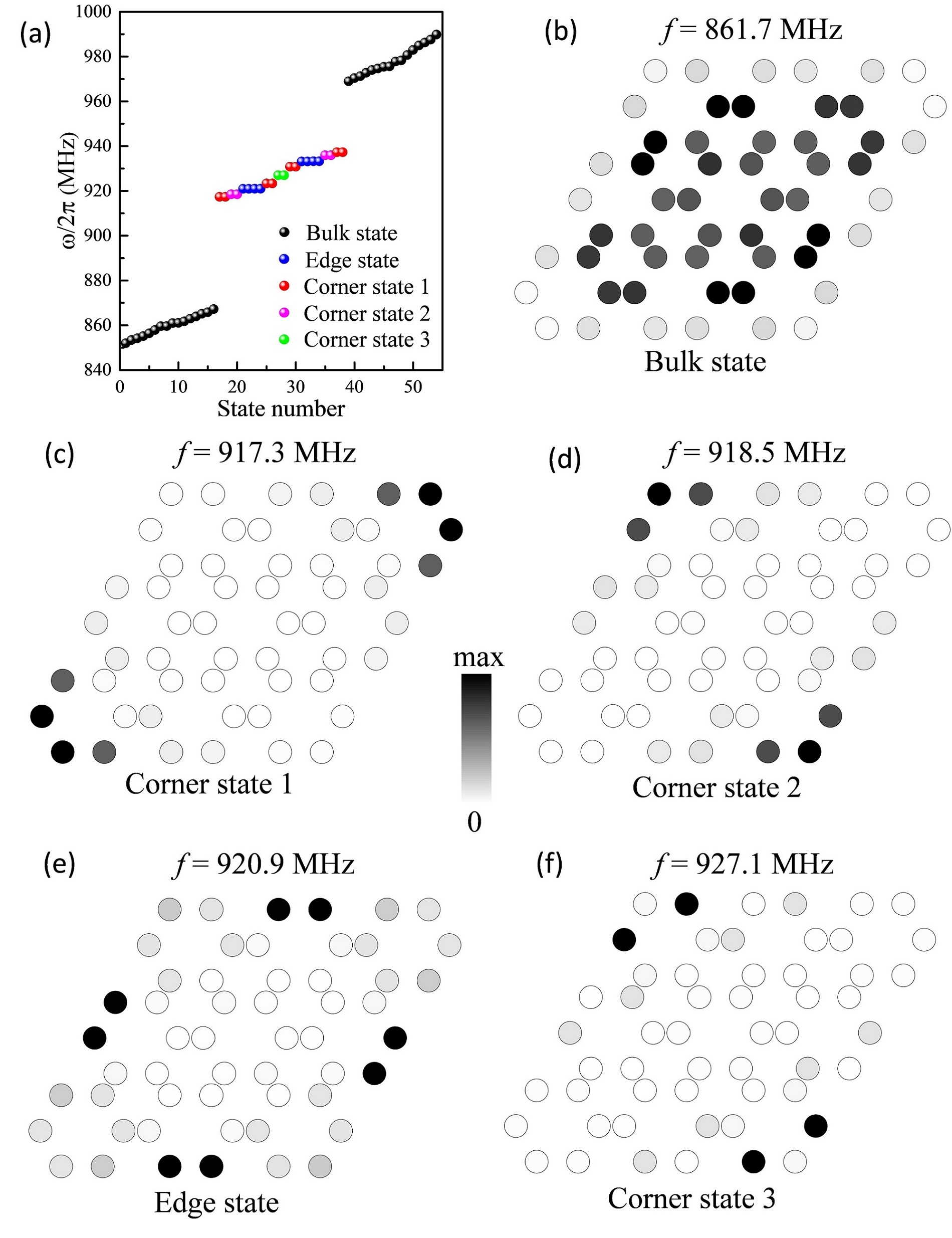}
\par\end{centering}
\caption{ (a) Eigenfrequencies of the finite system with parameters $d_{1}=2.08r$ and $d_{2}=3.6r$ for the parallelogram-shaped structure [see Fig. \ref{Figure3}(b)]. The spatial distribution of vortex gyrations for the bulk (b), corner (c, d and f), and edge (e) states of five representative frequencies.}
\label{Figure5}
\end{figure}

To visualize higher-order edge states or, more precisely, the second-order corner states, we consider a parallelogram-shaped vortex lattice, as shown in Fig. \ref{Figure3}(b), where $d_{1}=2.08r$ and $d_{2}=3.6r$ [corresponding to the red star in Fig. \ref{Figure3}(d)]. From the spectrum [see Fig. \ref{Figure5}(a)], we clearly see that there exist a few degenerate modes in the band gap. To distinguish these states, we plot the spatial distribution of vortex gyrations for each mode in Figs. \ref{Figure5}(b)-\ref{Figure5}(f). We confirm three types of corner states, all of which have oscillations highly localized at obtuse-angled or acute-angled corners [see Figs. \ref{Figure5}(c), \ref{Figure5}(d), and \ref{Figure5}(f)], where corner states 1 (type I), 2 (type II), and 3 (type III) are denoted by red, magenta, and green balls, respectively. Two degenerate edge modes are denoted by blue balls, in which only two vortices on each edge participate in the oscillation, as shown in Fig. \ref{Figure5}(e). Figure \ref{Figure5}(b) shows the bulk state with oscillations spreading over the whole lattice except the boundaries. To judge whether these edge and corner states are topologically protected or not, we introduce moderate defects and disorder into the system and evaluate the change in the spectrum. We find that the eigenfrequency of ``zero-energy" corner state 3 at the obtuse-angled corner [see Fig. \ref{Figure5}(f) and Fig. \ref{Figure7} in Appendix A] is well confined around 927 MHz, which means that this corner state is suitably immune from external frustrations. This feature is due to the topological protection from the generalized chiral symmetry, as argued above. However, the frequencies of other corner modes [Figs. \ref{Figure5}(c) and \ref{Figure5}(d)] have obvious shifts, revealing that these crystalline-symmetry-induced modes are sensitive to disorder. The origin of the edge state [Fig. \ref{Figure5}(e)] is attributed to the so-called Tamm-Shockley mechanism \cite{Tamm1932,Shockley1939}.

To figure out why the chiral symmetry-protected (CSP) corner modes emerge at only obtuse-angled corners instead of acute-angled corners, we introduce the topological index $\mathcal{N}=|\mathcal{N}_{+}-\mathcal{N}_{-}|$, which captures the interplay between the topology of the bulk Hamiltonian and the defect structure \cite{Teo2010}. Here, $\mathcal{N}$ counts the number of topologically stable modes bound to corners, and $\mathcal{N}_{\pm}$ are integers counting the number of eigenstates of the chiral symmetry operator $\hat{\Pi}$ with eigenvalues $+1$ and $-1$, respectively. In the zero-correlation length limit $d_{2} \rightarrow\infty$, the breathing honeycomb lattice is then reduced to isolated dimers [see Fig. \ref{Figure8}(b)]. As long as the gap is not closed, the symmetry and the Berry phase remain, as evidenced by the topological invariant $\theta$. When the system is in the HOTI phase, $\mathcal{N}_{+}=\mathcal{N}_{-}=1$ in each edge unit cell and $\mathcal{N}_{+}=\mathcal{N}_{-}=2$ for acute-angled corners, we thus have $\mathcal{N}=0$, indicating that there may exist non-CSP modes at acute-angled corners or edges. However, a similar analysis results in totally different outcomes for obtuse-angled corners: we have $\mathcal{N}_{+}=1$ and $\mathcal{N}_{-}=2$ or $\mathcal{N}_{+}=2$ and $\mathcal{N}_{-}=1$, which leads to $\mathcal{N}=1$. Thus, CSP or ``zero-energy" modes must exist in each obtuse-angled corner (see Appendix B for details).

We further study corner states in the same lattice but with a zigzag edge. Interestingly, we find that the ``zero-energy" corner state now appears at acute-angled corners rather than obtuse-angled corners, in sharp contrast to results with armchair edges. Again, this finding can be fully explained in terms of the topological index $\mathcal{N}$ (see Appendices B and C for details).

\subsection{Micromagnetic simulations}
\begin{figure}[ptbh]
\begin{centering}
\includegraphics[width=0.48\textwidth]{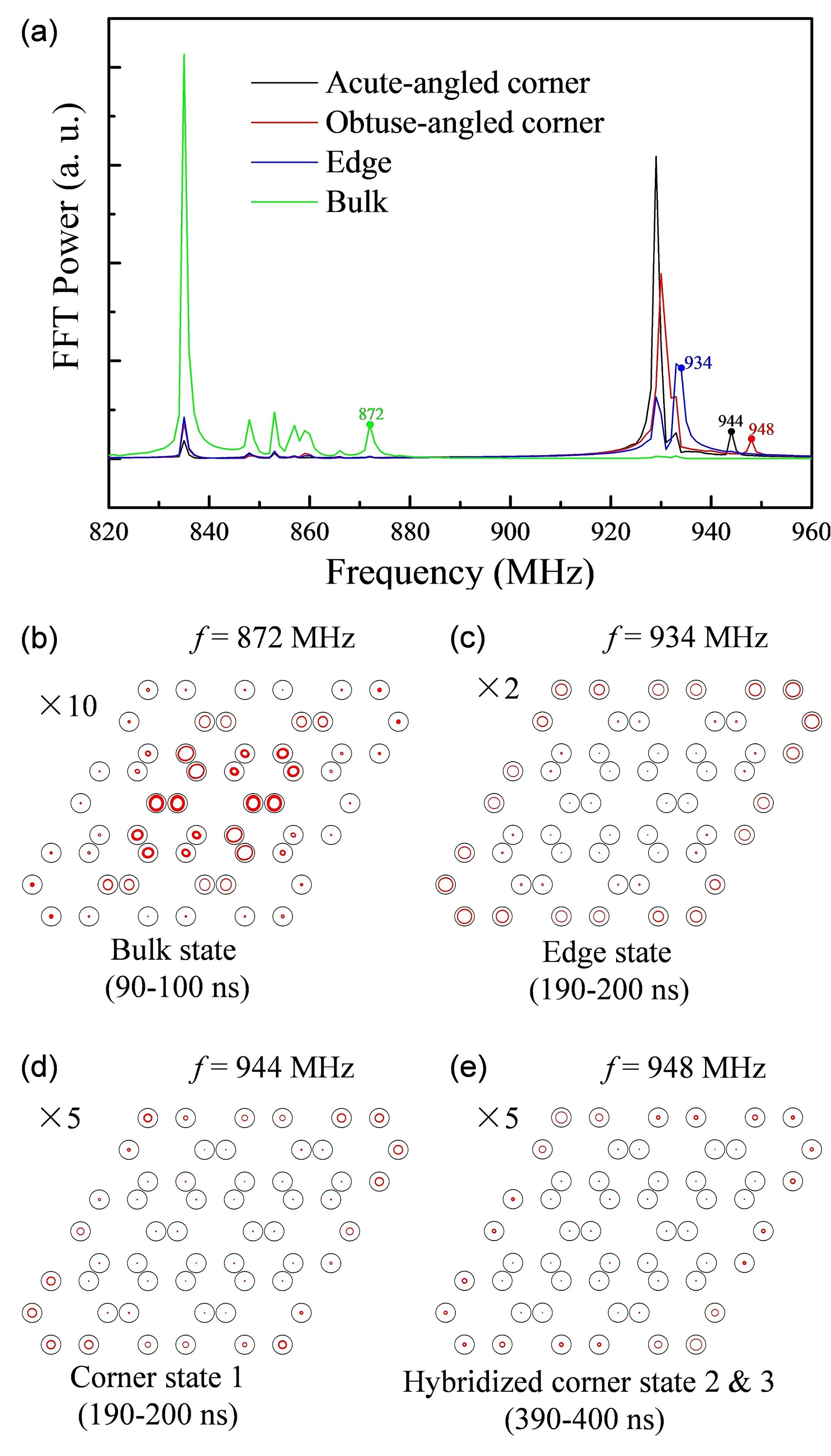}
\par\end{centering}
\caption{(a) The temporal Fourier spectra of the vortex oscillations at different positions marked by dashed black rectangles in Fig. \ref{Figure3}(b). The gyration path for all vortices under excitation fields with different frequencies, 872 MHz (b), 934 MHz (c), 944 MHz (d), and 948 MHz (e). Since the oscillation amplitudes of the vortex centers are too small, we have magnified them by 2, 5 or 10 times, as labeled in each figure.}
\label{Figure6}
\end{figure}
To verify our theoretical predictions, full micromagnetic simulations are implemented in the parallelogram-shaped breathing honeycomb lattice of magnetic vortices with an armchair edge, as shown in Fig. \ref{Figure3}(b). All material parameters adopted in the simulations are the same as those for the theoretical calculations in Fig. \ref{Figure4}(d): the saturation magnetization $M_{s}=0.86\times10^{6}$ A/m, the exchange stiffness $A=1.3\times10^{-11}$ J/m, and the Gilbert damping constant $\alpha=10^{-4}$ (choosing a small damping constant is to see the oscillation of vortices more clearly). The numerical package MUMAX3 \cite{Vansteenkiste2014} is used to simulate the collective dynamics of vortices in the artificial lattice. The mesh size is set to $2\times2\times10 $ nm$^{3}$. To obtain the excitation spectrum of the collective oscillation of vortices, the sinc-function magnetic field $H(t)=H_{0}\sin[2\pi$\emph{f}$(t-t_{0})]/[2\pi$\emph{f}$(t-t_{0})]$ along the $x$-direction with $H_{0}=10$ mT, $f=10$ GHz, and $t_{0}=1$ ns is applied to the whole system for 1 $\mu$s, which gives a very high frequency-resolution of 1 MHz. The vortex centers $\textbf{R}_{j}=(R_{j,x}, R_{j,y}$) in all nanodisks are recorded every 200 ps. Here, $R_{j,x}=\frac {\int \!\!\! \int{x|m_{z}|^{2}dxdy}}{\int \!\!\! \int{|m_{z}|^{2}dxdy}}$, and $R_{j,y}=\frac {\int \!\!\! \int{y|m_{z}|^{2}dxdy}}{\int \!\!\! \int{|m_{z}|^{2}dxdy}}$, with the integral region confined in the $j$-th nanodisk.

To identify the emerging corner states, edge states, and bulk states shown in Fig. \ref{Figure5}, we compute the temporal Fourier spectrum of the vortex oscillations at different positions [labeled with dashed black rectangles in Fig. \ref{Figure3}(b)]. Figure \ref{Figure6}(a) shows the spectra, with the black, red, blue, and green curves indicating the positions of acute-angled corner, obtuse-angled corner, and edge and bulk bands, respectively. We can clearly see that around the frequency of 944 MHz (948 MHz), the spectra for acute-angled corner (obtuse-angled corner) have an obvious peak, which does not happen for the spectra for edge and bulk bands. Therefore, we believe that these two peaks denote two different corner states that are located at acute-angled or obtuse-angled corners. Similarly, we can identify the frequency range for bulk and edge states. Further, to visualize the spatial distribution of the vortex oscillations for different modes, four representative frequencies are chosen and are marked by green, blue, black and red dots: 872 MHz for the bulk state, 934 MHz for the edge state, 944 MHz for the acute-angled corner state, and 948 MHz for the obtuse-angled corner state, respectively. We then stimulate their dynamics by applying a sinusoidal field $\textbf{h}(t)=h_0\sin(2\pi ft)\hat{x}$ with $h_0=0.1$ mT applied to the whole system. The 10 ns gyration paths of all vortices are plotted in Figs. \ref{Figure6}(b)-\ref{Figure6}(e) after the excitation field drives steady-state vortex dynamics. The spatial distribution of vortices motion for the bulk and edge states are shown in Fig. \ref{Figure6}(b) and Fig. \ref{Figure6}(c), respectively. We also observe the type I corner state with vortex oscillation localized at the acute-angled corner in Fig. \ref{Figure6}(d), which is in good agreement with the theoretical result. Interestingly, we note strong hybridization between the type II and type III corner states, as shown in Fig. \ref{Figure5}(f), which is because their frequencies are very close to each other [see Fig. \ref{Figure5}(a)] and their wavefunctions have a large overlap [see Figs. \ref{Figure5}(d) and \ref{Figure5}(f)].

\section{CONCLUSION}
To conclude, we have investigated the topological phases of a breathing honeycomb lattice of magnetic vortices with both armchair and zigzag edges. We generalized the conventional chiral symmetry to sexpartite lattices with $C_{6}$ symmetry. The phase diagram was obtained theoretically by computing the Chern number and $\mathbb{Z}_{6}$ Berry phase, the quantization of which enabled us to identify three different phases allowed in the system: the trivial, FOTI and HOTI phases. We showed that the shape of the lattice boundary determines the position of corner states: CSP corner states always exist at obtuse-angled $2\pi/3$ corners for armchair edges and at acute-angled $\pi/3$ corners for zigzag edges. We note that both fabricating the metamaterials of present interest and detecting the highly spatially localized corner mode are already within the reach of current technology. We thus strongly believe that our findings will provide important theoretical reference for fully understanding topological phases and their transition in magnetic systems and for finally achieving spin texture-based higher-order modes with practical merit in robust spintronic information processing by harvesting their topological features. The interaction effect in HOTIs is also an interesting issue for future study.

\begin{acknowledgments}
\section*{ACKNOWLEDGMENTS}
We thank C. Wang, X. S. Wang, and Z. Wang for helpful discussions. This work was supported by the National Natural Science Foundation of China (NSFC) (Grants No. 11604041 and 11704060), the National Key Research Development Program under Contract No. 2016YFA0300801, and the National Thousand-Young-Talent Program of China. X. R. Wang was supported by Hong Kong RGC (Grants No. 16300117, 16301518, and 16301619). Z.-X. Li acknowledges the financial support of NSFC Grant No. 11904048.
\end{acknowledgments}

\section*{APPENDIX A: ROBUSTNESS OF CORNER STATES}
\begin{figure}[ptbh]
\begin{centering}
\includegraphics[width=0.48\textwidth]{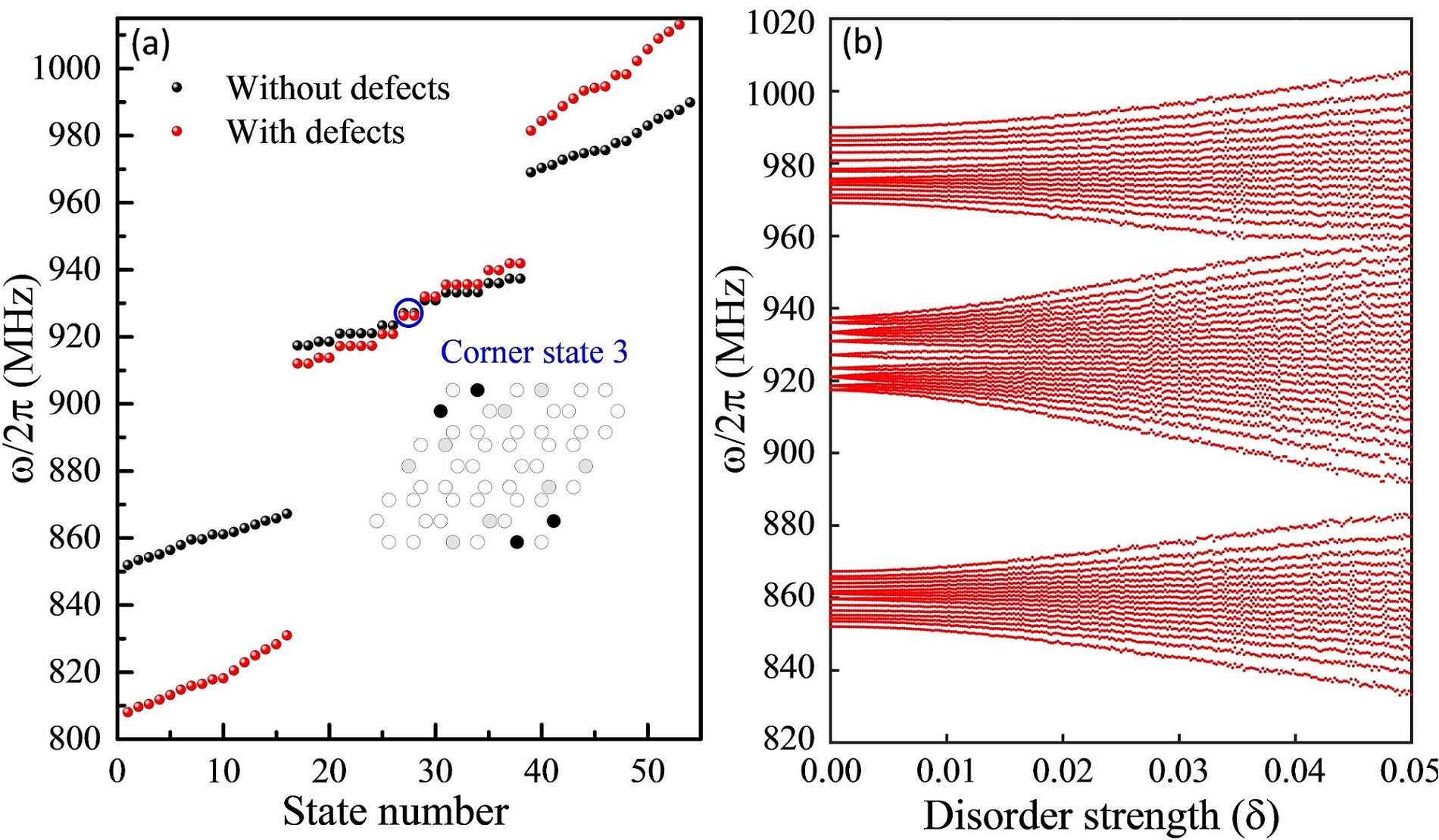}
\par\end{centering}
\caption{(a) Eigenfrequencies of the parallelogram-shaped vortex lattice with armchair edges in the absence of (black balls) and in the presence of defects (red balls). The blue circle indicates the topologically stable corner state 3 (type III), with the inset plotting the spatial distribution of vortex gyrations. (b) Eigenfrequencies of the parallelogram-shaped vortex lattice under different disorder strengths. In the calculations, we set the alternating bond lengths $d_{1}=2.08r$ and $d_{2}=3.6r$.}
\label{Figure7}
\end{figure}

Topologically protected edge states or corner states should be robust under moderate disorder and defects. To judge whether the edge and corner states emerging in Fig. \ref{Figure5} have a topological nature, we study the vortex gyration spectrum by creating some moderate defects or disorder, with the computed results plotted in Figs. \ref{Figure7}(a) and \ref{Figure7}(b), respectively. Here, defects are introduced by assuming that coupling parameters $\zeta$ and $\xi$ undergo a shift ($\zeta \rightarrow 1.5\zeta$, $\xi \rightarrow 1.5\xi$). Moreover, disorder is introduced by assuming that the resonant frequency $\omega_{0}$ undergo a random shift, i.e., $\omega_{0} \rightarrow \omega_{0}+\delta Z\omega_{0}$, where $\delta$ measures the strength of the disorder and $Z$ is a uniformly distributed random number between $-1$ to $1$. We average the calculation after 100 realizations. In Fig. \ref{Figure7}(a), we clearly see that the frequency of corner state 3 is well confined at approximately 927 MHz. Nevertheless, frequencies of other edge and corner states acquire an obvious deviation from the clean limit (black balls). Further, with increasing of disorder strength [see Fig. \ref{Figure7}(b)], the spectra of the edge state, corner state 1 and corner state 2 are significantly modified and even shift to the spectrum of the bulk band when disorder strength exceeds a critical value, while corner state 3 is quite robust, although the frequencies have changed to some extent. We thus verify that the ``zero-energy" mode protected by the generalized chiral symmetry is much more robust than the ones protected by the crystalline symmetry.

\begin{figure}[ptbh]
\begin{centering}
\includegraphics[width=0.48\textwidth]{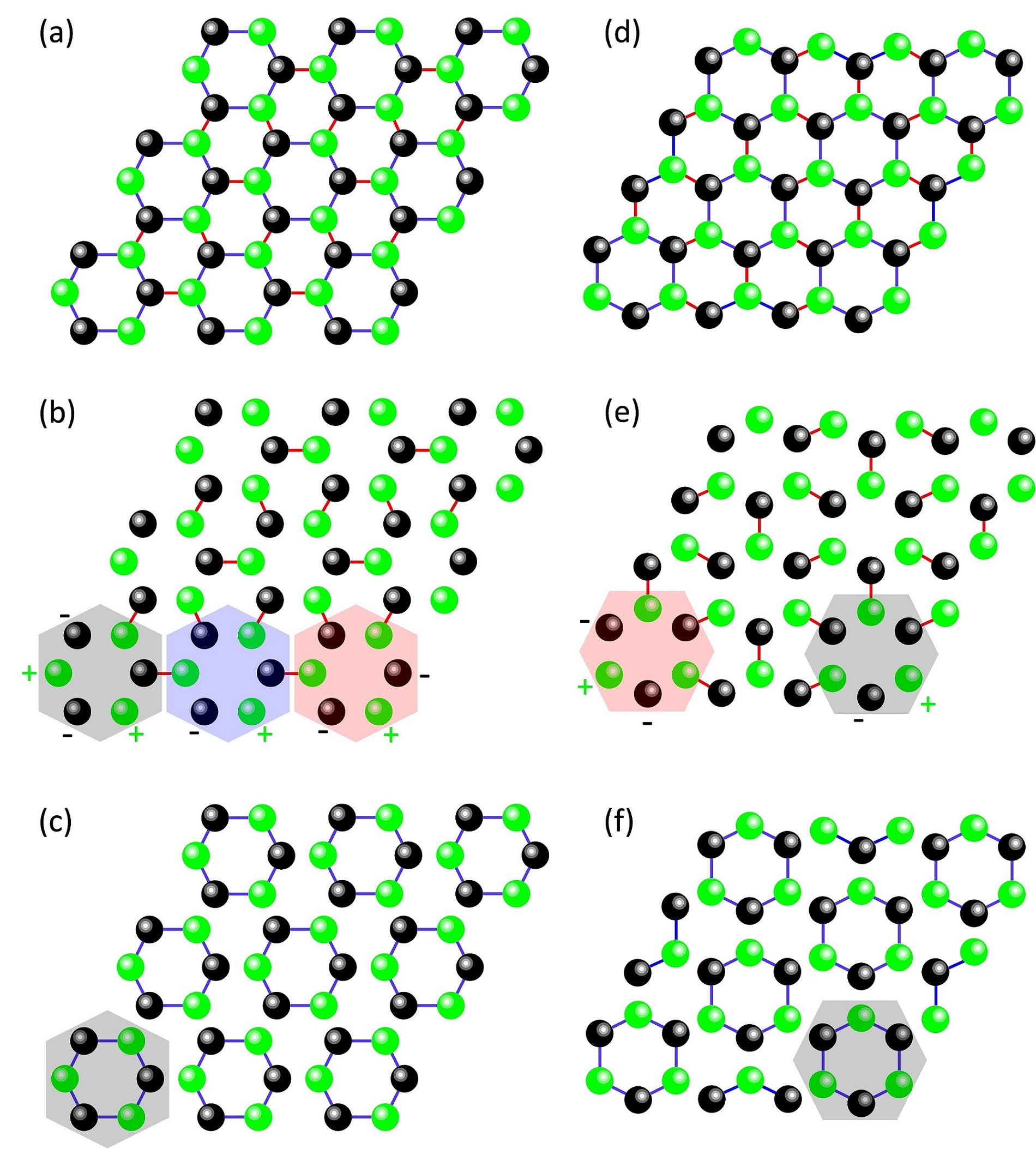}
\par\end{centering}
\caption{The configuration of breathing honeycomb lattices of magnetic vortices with armchair (a) and zigzag edges (d). (b) and (e) [(c) and (f)] are the corresponding configurations of (a) and (d) in the zero-correlation length limit $d_{2} \rightarrow\infty$ ($d_{1}\rightarrow\infty$), respectively, which consist of isolated dimers (hexamer). Green and black balls indicate eigenvalues of $+1$ and $-1$ of the chiral-symmetry operator, respectively. Shaded areas represent the unit cell at different positions.}
\label{Figure8}
\end{figure}
\section*{APPENDIX B: Topological index $\mathcal{N}$ for determining the position of corner modes}
The location of topological corner states in the finite lattice can be well explained by the topological index $\mathcal{N}$. When the system is in the HOTI phase, parallelogram-shaped vortex lattices with two different edges, i.e., armchair and zigzag, are considered; see Figs. \ref{Figure8}(a) and \ref{Figure8}(d). We plot their configuration in the zero-correlation length limit, i.e., the intracellular (intercellular) bond length $d_{2} \rightarrow\infty$ ($d_{1}\rightarrow\infty$) such that the coupling within (between) the unit cell approaches zero, in Figs. \ref{Figure8}(b) and \ref{Figure8}(e) [Figs. \ref{Figure8}(c) and \ref{Figure8}(f)], respectively. The phase diagram [see Fig. \ref{Figure3}(d)] indicates that configurations (b) and (e) [(c) and (f)] are in the HOTI (trivial) phase. From Fig. \ref{Figure8} (b), we can clearly see that there are four modes at the acute-angled corner unit cell (black shadow area) and that two of them have eigenvalues of $+1$, whereas the other two have eigenvalues of $-1$; thus, $\mathcal{N}=|\mathcal{N}_{+}-\mathcal{N}_{-}|=|2-2|=0$. Similarly, we obtain $\mathcal{N}=0$ for the edge unit cell and $\mathcal{N}=1$ for the obtuse-angled corner unit cell. Therefore, for the configuration with armchair edges, there are CSP stable corner states existing at only obtuse-angled corners. However, for the case with zigzag edges [see Fig. \ref{Figure8}(e)], we obtain $\mathcal{N}=1$ for the acute-angled corner and $\mathcal{N}=0$ for the obtuse-angled corner, indicating that CSP stable corner states appear at acute-angled corners instead of obtuse-angled corners. On the other hand, in the limit $d_{1}\rightarrow\infty$ [see Figs. \ref{Figure8}(c) and \ref{Figure8}(f)], we find that there are no uncoupled vortices. The six corners of the isolated hexagon are equivalent, with no special edge or corner states.

\begin{figure}[ptbh]
\begin{centering}
\includegraphics[width=0.48\textwidth]{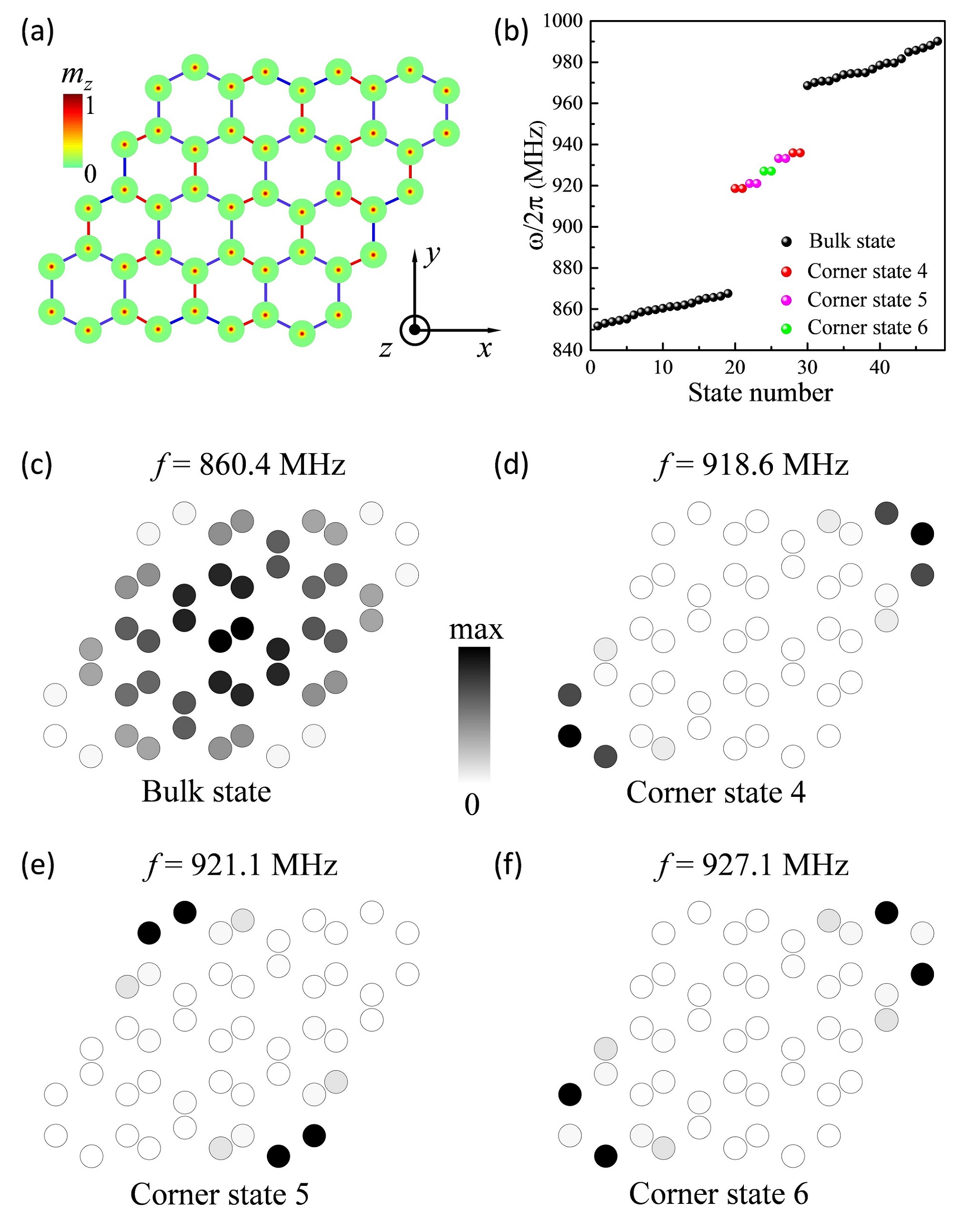}
\par\end{centering}
\caption{(a) Schematic plot of the parallelogram-shaped vortex lattice with zigzag edges. (b) Eigenfrequencies with the parameters $d_{1}=2.08r$ and $d_{2}=3.6r$. The spatial distribution of vortex gyrations for the bulk (c) and corner (d, e and f) states with four representative frequencies.}
\label{Figure9}
\end{figure}

\section*{APPENDIX C: Corner states in zigzag-edge lattice}
For completeness, we present detailed wavefunction calculations to study corner states in a breathing honeycomb lattice of vortices with zigzag edges, with the sketch plotted in Fig. \ref{Figure9}(a). The same parameters as those in the armchair configuration ($d_{1}=2.08r, d_{2}=3.6r$) are considered. Figure \ref{Figure9}(b) shows the eigenspectrum of the system, obtained by numerically solving Eq. \eqref{Eq2} or Eq. \eqref{Eq4}. By drawing the spatial distribution of vortex oscillation for different modes [see Figs. \ref{Figure9}(c)-\ref{Figure9}(f)], we confirm three different corner states, which are labeled 4, 5, and 6, and represented by the red, magenta and green balls, respectively; see Fig. \ref{Figure9}(b). In contrast to the bulk states plotted in Fig. \ref{Figure9}(c), the modes shown in Figs. \ref{Figure9}(d)-\ref{Figure9}(f) exhibit the characteristics of local oscillations.
\begin{figure}[ptbh]
\begin{centering}
\includegraphics[width=0.48\textwidth]{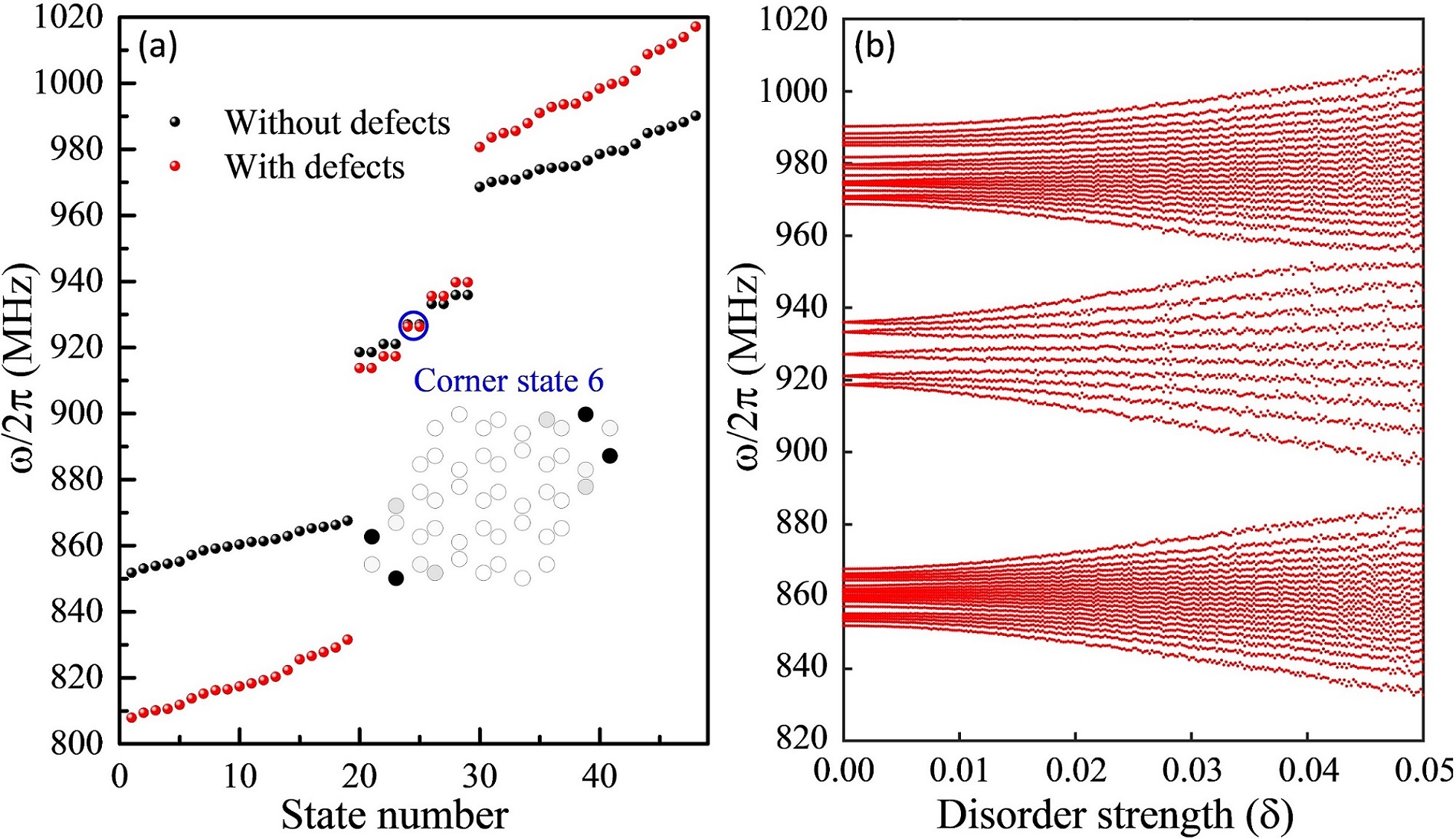}
\par\end{centering}
\caption{(a) Eigenfrequencies of the parallelogram-shaped vortex with zigzag edges lattice in the absence of (black balls) and in the presence of defects (red balls). The blue circle indicates corner state 6 (type III), with the inset plotting the corresponding spatial distribution of vortex gyrations. (b) Eigenfrequencies of the system under different disorder strengths. In the calculations, we set the same parameters as those used in Fig. \ref{Figure7}.}
\label{Figure10}
\end{figure}

Furthermore, we also investigate the topological stability of these corner states by introducing defects and disorder. Figure \ref{Figure10}(a) shows the eigenfrequencies of the system with defects (red balls) and without defects (black balls). The effect of disorder on eigenfrequencies is shown in Fig. \ref{Figure10}(b). Similar to the lattice with armchair edges, we found that corner state 6 is quite robust and is protected by the generalized chiral symmetry, while the other two corner states 4 and 5 undergo an obvious frequency shift. These results can also be explained by the topological index $\mathcal{N}$, as discussed in Appendix B.

\end{document}